\documentclass[12pt]{article}
\usepackage{graphicx}
\def\bra#1{\left\langle #1\right|}
\def\ket#1{\left| #1\right\rangle}
\newcommand{\bers}{\begin{eqnarray*}}
\newcommand{\eers}{\end{eqnarray*}}
\newcommand{\bt}{\begin{itemize}}
\newcommand{\et}{\end{itemize}}
\def\beq{\begin{equation}}
\def\eeq{\end{equation}}
\def\bea{\begin{eqnarray}}
\def\eea{\end{eqnarray}}
\def\nn{\nonumber}
\def\sss{\scriptscriptstyle}
\def\bd{B_d^0}
\def\bdbar{{\overline{B_d^0}}}
\def\bs{B_s^0}
\def\bsbar{{\overline{B_s^0}}}
\def\barp{{\raise.35ex\hbox
{${\sss (}$}}---{\raise.35ex\hbox{${\sss )}$}}}
\def\bdbarp{\hbox{$B_d$\kern-1.4em\raise1.4ex\hbox{\barp}}}
\def\bsbarp{\hbox{$B_s$\kern-1.4em\raise1.4ex\hbox{\barp}}}
\def\ks{K_{\sss S}}
\def\kbar{{\overline{K^0}}}
\def\roughly#1{\mathrel{\raise.3ex\hbox
{$#1$\kern-.75em\lower1ex\hbox{$\sim$}}}}

\def\gsim{\roughly>}
\def\mK{m_{\sss K}}
\def\epjc#1#2#3{{\it Eur.\ Phys.\ J.}\ {\bf C#1} (#2) #3}

\def\jhep#1#2#3{{\it JHEP} {\bf B#1} (#2) #3}

\def\npb#1#2#3{{\it Nucl.\ Phys.} {\bf B#1} (#2) #3}
\def\plb#1#2#3{{\it Phys.\ Lett.} {\bf B#1} (#2) #3}

\def\prd#1#2#3{{\it Phys.\ Rev.} {\bf D#1} (#2) #3}
\def\newprd#1#2#3{{\it Phys.\ Rev.} {\bf D#1}: #3 (#2)}

\def\prl#1#2#3{{\it Phys.\ Rev.\ Lett.} {\bf #1} (#2) #3}

\def\zpc#1#2#3{{\it Zeit.\ Phys.} {\bf C#1} (#2) #3}

\newread\epsffilein 
\newif\ifepsffileok 
\newif\ifepsfbbfound 
\newif\ifepsfverbose 
\newdimen\epsfxsize 
\newdimen\epsfysize 
\newdimen\epsftsize 
\newdimen\epsfrsize 
\newdimen\epsftmp 
\newdimen\pspoints 
\def\epsfbox#1{\global\def\epsfllx{72}\global\def\epsflly{72}%
 \global\def\epsfurx{540}\global\def\epsfury{720}%
 \def\lbracket{[}\def\testit{#1}\ifx\testit\lbracket
 \let\next=\epsfgetlitbb\else\let\next=\epsfnormal\fi\next{#1}}%
\def\epsfgetlitbb#1#2 #3 #4 #5]#6{\epsfgrab #2 #3 #4 #5 .\\%
 \epsfsetgraph{#6}}%
\def\epsfnormal#1{\epsfgetbb{#1}\epsfsetgraph{#1}}%
\def\epsfgetbb#1{%
\openin\epsffilein=#1
\ifeof\epsffilein\errmessage{I couldn't open #1, will ignore it}\else
 {\epsffileoktrue \chardef\other=12
 \def\do##1{\catcode`##1=\other}\dospecials \catcode`\ =10
 \loop
 \read\epsffilein to \epsffileline
 \ifeof\epsffilein\epsffileokfalse\else
 \expandafter\epsfaux\epsffileline:. \\%
 \fi
 \ifepsffileok\repeat
 \ifepsfbbfound\else
 \ifepsfverbose\message{No bounding box comment in #1; using defaults}\fi\fi
 }\closein\epsffilein\fi}%
\def\epsfclipstring{}
\def\epsfsetgraph#1{%
 \epsfrsize=\epsfury\pspoints
 \advance\epsfrsize by-\epsflly\pspoints
 \epsftsize=\epsfurx\pspoints
 \advance\epsftsize by-\epsfllx\pspoints
 \epsfxsize\epsfsize\epsftsize\epsfrsize
 \ifnum\epsfxsize=0 \ifnum\epsfysize=0
 \epsfxsize=\epsftsize \epsfysize=\epsfrsize
 \epsfrsize=0pt
 \else\epsftmp=\epsftsize \divide\epsftmp\epsfrsize
 \epsfxsize=\epsfysize \multiply\epsfxsize\epsftmp
 \multiply\epsftmp\epsfrsize \advance\epsftsize-\epsftmp
 \epsftmp=\epsfysize
 \loop \advance\epsftsize\epsftsize \divide\epsftmp 2
 \ifnum\epsftmp>0
 \ifnum\epsftsize<\epsfrsize\else
 \advance\epsftsize-\epsfrsize \advance\epsfxsize\epsftmp \fi
 \repeat
 \epsfrsize=0pt
 \fi
 \else \ifnum\epsfysize=0
 \epsftmp=\epsfrsize \divide\epsftmp\epsftsize
 \epsfysize=\epsfxsize \multiply\epsfysize\epsftmp
 \multiply\epsftmp\epsftsize \advance\epsfrsize-\epsftmp
 \epsftmp=\epsfxsize
 \loop \advance\epsfrsize\epsfrsize \divide\epsftmp 2
 \ifnum\epsftmp>0
 \ifnum\epsfrsize<\epsftsize\else
 \advance\epsfrsize-\epsftsize \advance\epsfysize\epsftmp \fi
 \repeat
 \epsfrsize=0pt
 \else
 \epsfrsize=\epsfysize
 \fi
 \fi
 \ifepsfverbose\message{#1: width=\the\epsfxsize, height=\the\epsfysize}\fi
 \epsftmp=10\epsfxsize \divide\epsftmp\pspoints
 \vbox to\epsfysize{\vfil\hbox to\epsfxsize{%
 \ifnum\epsfrsize=0\relax
 \includegraphics{#1}%
 \else
 \epsfrsize=10\epsfysize \divide\epsfrsize\pspoints
 \includegraphics{#1}%
 \fi
 \hfil}}%
\global\epsfxsize=0pt\global\epsfysize=0pt}%
 {\catcode`\%=12 \global\let\epsfpercent=
\long\def\epsfaux#1#2:#3\\{\ifx#1\epsfpercent
 \def\testit{#2}\ifx\testit\epsfbblit
 \epsfgrab #3 . . . \\%
 \epsffileokfalse
 \global\epsfbbfoundtrue
 \fi\else\ifx#1\par\else\epsffileokfalse\fi\fi}%
\def\epsfempty{}%
\def\epsfgrab #1 #2 #3 #4 #5\\{%
\global\def\epsfllx{#1}\ifx\epsfllx\epsfempty
 \epsfgrab #2 #3 #4 #5 .\\\else
 \global\def\epsflly{#2}%
 \global\def\epsfurx{#3}\global\def\epsfury{#4}\fi}%
\def\epsfsize#1#2{\epsfxsize}
\pagestyle{plain}

\begin{document}

\begin{flushright}  
UdeM-GPP-TH-01-84\\
\end{flushright}

\begin{center} 

{\large \bf
\centerline{Measuring $\sin 2\beta$ in $\bs(t) \to \phi \ks$}}
\vspace*{1.0cm}
{\large Alakabha Datta$^{a,}$\footnote{email: datta@lps.umontreal.ca},
  C.S. Kim$^{b,}$\footnote{email: kim@kimcs.yonsei.ac.kr,
    cskim@pheno.physics.wisc.edu} and David
  London$^{a,}$\footnote{email: london@lps.umontreal.ca}} \vskip0.3cm
{\it ${}^a$ Laboratoire Ren\'e J.-A. L\'evesque, Universit\'e de
  Montr\'eal,} \\
{\it C.P. 6128, succ.\ centre-ville, Montr\'eal, QC, Canada H3C 3J7} \\
\vskip0.3cm
{\it ${}^b$ Department of Physics, Yonsei University, Seoul 120-749, Korea} \\
\vskip0.5cm
\bigskip
(\today)
\vskip0.5cm
{\Large Abstract\\}
\vskip3truemm
\parbox[t]{\textwidth} { We show that, unlike other pure $b\to d$
  penguin processes, the decay $\bs(t)\to\phi\ks$ is dominated by a
  single amplitude, that of the internal $t$-quark. The contributions
  of the $u$- and $c$-quark operators each vanish due to a
  cancellation between the $(V-A) \otimes (V-A)$ and $(V-A) \otimes
  (V+A)$ matrix elements. Thus, the indirect CP asymmetry in this
  decay probes $\sin 2\beta$. Although this cancellation is complete
  only for certain values of the $s$- and $b$-quark masses, the
  theoretical uncertainty on $\sin 2\beta$ is still less than 10\%
  over most ($\sim 80$\%) of the parameter space. By measuring the
  direct CP asymmetry, one can get a better idea of the probable error
  on $\sin 2\beta$.}
\end{center}
\thispagestyle{empty}
\newpage
\setcounter{page}{1}
\textheight 23.0 true cm
\baselineskip=14pt

It has been known for many years that the $B$ system is a particularly
good place to test the standard model (SM) explanation of CP
violation. By measuring CP-violating rate asymmetries in the decays of
neutral $B$ mesons to a variety of final states, one can cleanly
extract the CP phases $\alpha$, $\beta$ and $\gamma$ \cite{BCPreview}.
This allows one to construct the unitarity triangle \cite{PDG} and
search for the presence of physics beyond the SM.

In the early days of the field, only tree-level decays of $B$ mesons
were considered. However, it was soon realized that penguin amplitudes
could play an important role \cite{LonPeccei,penguins}. For example,
the presence of penguin contributions in $\bd(t) \to \pi^+\pi^-$ can
spoil the clean extraction of $\alpha$ (though this can be rectified
with the help of an isospin analysis \cite{isospin}). And the clean
measurement of $\gamma$ via $\bs(t) \to \rho\ks$ is completely ruined
since, for this decay, the penguin amplitude is the dominant
contribution.

Given the importance of such penguin contributions, one is immediately
led to consider CP violation in pure penguin decays. In the
(approximate) Wolfenstein parametrization \cite{Wolfenstein} of the
Cabibbo-Kobayashi-Maskawa (CKM) matrix, there are only two matrix
elements which have a nonzero weak phase: $V_{td} \propto
\exp(-i\beta)$ and $V_{ub} \propto \exp(-i\gamma)$. Thus, assuming
that the penguin amplitudes are dominated by an internal $t$-quark,
one expects that the $b\to s$ penguin amplitude, which involves the
product of CKM matrix elements $V_{tb} V_{ts}^*$, is real, to a good
approximation. Similarly, the weak phase of the $b\to d$ penguin
amplitude ($V_{tb}V_{td}^*$) is $+\beta$. Knowing that the weak phases
of $\bd$-$\bdbar$ and $\bs$-$\bsbar$ mixing are, respectively,
$-\beta$ and 0, this allows us to compute the weak phase probed in
various pure-penguin decay asymmetries \cite{LonPeccei}:
\bea
b \to d & : & {\rm Asym}(\bd(t) \to K^0\kbar) \sim 0 ~, 
\label{Bdbtod} \\
        & & {\rm Asym}(\bs(t) \to \phi \ks) \sim -\sin 2\beta ~, 
\label{Bsbtod} \\
b \to s & : & {\rm Asym}(\bd(t) \to \phi \ks) \sim +\sin 2\beta ~, 
\label{Bdbtos} \\
        & & {\rm Asym}(\bs(t) \to \phi \phi) \sim 0 ~.
\label{Bsbtos} 
\eea

The problem with the above analysis is that the $b\to d$ penguin
amplitude is {\it not} dominated by an internal $t$-quark. In the
quark-level decays $b\to u{\bar u} d$ and $b \to c{\bar c} d$, the
$u{\bar u}$ and $c{\bar c}$ quark pairs can rescatter strongly into an
$s {\bar s}$ quark pair, giving effective $V_{ub} V_{ud}^*$ and
$V_{cb} V_{cd}^*$ contributions to the $b\to d$ penguin decays above.
Buras and Fleischer have estimated that these contributions can be
between 20\% and 50\% of the leading $t$-quark contribution
\cite{ucquark}. And since the $u$- and $c$-quark contributions have a
different weak phase than that of the $t$-quark contribution, this
implies that the weak phase of the $b\to d$ penguin amplitude is {\it
not} $+\beta$, so that the predictions of Eqs.~(\ref{Bdbtod}) and
(\ref{Bsbtod}) are not valid. On the contrary, due to the presence of
these several decay amplitudes, one expects that a weak phase {\it
cannot} be cleanly extracted from the measurement of CP asymmetries in
pure $b\to d$ penguin decays. One also expects to observe direct CP
violation in such decays.

Note that these same conclusions do not hold for $b\to s$ penguin
amplitudes. In this case, the CKM matrix-element product associated
with the $u$-quark contribution, $V_{ub} V_{us}^*$, is a small
fraction ($\sim 2\%$) of those of the corresponding $c$- and $t$-quark
contributions, $V_{cb} V_{cs}^*$ and $V_{tb} V_{ts}^*$. Since the $c$-
and $t$-quark CKM matrix elements are both real, the assumption that
the $b\to s$ penguin amplitude is real is a good approximation. Thus,
the predictions of Eqs.~(\ref{Bdbtos}) and (\ref{Bsbtos}) still hold.

In this paper, we re-examine the question of the weak phase of the
$b\to d$ penguin for the exclusive decay $\bs(t) \to \phi\ks$
[Eq.~(\ref{Bsbtod})]. As we will show, although the quark-level
contributions from $u$- and $c$-quarks are non-negligible, at the
meson level the matrix elements involving the corresponding $u$- and
$c$-quark operators each vanish, to a good approximation, over a large
region of parameter space. This is due to two factors. First, for a
large range of values of the $s$- and $b$-quark masses, it is a
fortuitous numerical coincidence that the matrix element of the $(V-A)
\otimes (V-A)$ piece of the $u$/$c$-quark operator is approximately
equal to that of the $(V-A) \otimes (V+A)$ piece. Second, because the
final state consists of a vector meson and a pseudoscalar meson, the
full matrix element is proportional to the difference between these
two pieces. In other words, there is a cancellation between the $(V-A)
\otimes (V-A)$ and $(V-A) \otimes (V+A)$ matrix elements. Thus, to the
extent that this cancellation is complete, the CP-violating rate
asymmetry in $\bs(t) \to \phi\ks$ still cleanly probes the weak phase
$\beta$.

We begin the discussion by considering the SM effective hamiltonian
for hadronic $B$ decays of the type $b\to d \bar{f} f$ \cite{Reina}:
\begin{eqnarray}
H_{eff}^{b\to d} &=& {G_F \over \protect \sqrt{2}}
[V_{fb}V^*_{fd}(c_1O_{1f}^d + c_2 O_{2f}^d) \nn\\ 
& & \qquad\qquad - \sum_{i=3}^{10}(V_{ub}V^*_{ud} c_i^u
+V_{cb}V^*_{cd} c_i^c +V_{tb}V^*_{td} c_i^t) O_i^d] + h.c.
\end{eqnarray}
In the first two terms, $f$ can be a $u$ or a $c$ quark, while in the
last three terms, the superscript $u$, $c$ or $t$ indicates the
flavour of the internal quark. The operators $O_i^d$ are defined as
\begin{eqnarray}
O_{1f}^d = \bar d_\alpha \gamma_\mu Lf_\beta\bar 
f_\beta\gamma^\mu Lb_\alpha &~~~~~~&
O_{2f}^d =\bar d \gamma_\mu L f\bar f\gamma^\mu L b \nn\\
O_{3,5}^d =\bar d \gamma_\mu L b \bar q' \gamma_\mu L(R) q'
&~~~~~~&
O_{4,6}^d = \bar d_\alpha \gamma_\mu Lb_\beta 
\bar q'_\beta \gamma_\mu L(R) q'_\alpha \\
O_{7,9}^d = {3\over 2}\bar d \gamma_\mu L b e_{q'}\bar q'
\gamma^\mu R(L)q' &~~~~~~&
O_{8,10}^d = {3\over 2}\bar d_\alpha \gamma_\mu L b_\beta e_{q'}\bar
q'_\beta \gamma_\mu R(L) q'_\alpha ~, \nn
\end{eqnarray}
where $R(L) = 1 \pm \gamma_5$, and $q'$ is summed over $u$, $d$ and
$s$. $O_1$ and $O_2$ are, respectively, the direct and QCD-corrected
tree-level operators, $O_{3-6}$ are the strong gluon-induced penguin
operators, and operators $O_{7-10}$ are due to $\gamma$ and $Z$
exchange (electroweak penguins), and ``box'' diagrams at loop level.
The Wilson coefficients $c_i^f$ are defined at the scale $\mu \approx
m_b$ and have been evaluated to next-to-leading order in QCD. Below we
give the regularization-scheme-independent values for the $c_i^f$ for
$m_t = 176$ GeV, $\alpha_s(m_Z) = 0.117$, and $\mu = m_b = 5$ GeV
\cite{FSHe}:
\begin{eqnarray}
c_1 = -0.307 &,& c_2 = 1.147 \nn\\
c^t_3 = 0.017 ~,~~ c^t_4 = -0.037 &,& c^t_5 = 0.010 ~,~~
c^t_6 =-0.045 ~, \nn\\
c^t_7 = -1.24\times 10^{-5} ~,~~ c_8^t = 3.77\times 10^{-4} &,&
c_9^t = -0.010 ~,~~ c_{10}^t = 2.06\times 10^{-3} ~, \nn\\
c_{3,5}^i = -c_{4,6}^i/N_c = P^i_s/N_c &,&
c_{7,9}^i = P^i_e ~,~~
c_{8,10}^i = 0 ~,~~
i=u,c ~,
\label{coeffs}
\end{eqnarray}
where $N_c$ is the number of colors. The leading contributions to
$P^i_{s,e}$ are given by $P^i_s = ({\frac{\alpha_s}{8\pi}}) c_2
({\frac{10}{9}} +G(m_i,\mu,q^2))$ and $P^i_e =
({\frac{\alpha_{em}}{9\pi}}) (N_c c_1+ c_2) ({\frac{10}{9}} +
G(m_i,\mu,q^2))$, where the function $G(m,\mu,q^2)$ takes the form
\begin{eqnarray}
G(m,\mu,q^2) = 4\int^1_0 x(1-x) \mbox{ln}{m^2-x(1-x)q^2\over
\mu^2} ~\mbox{d}x ~.
\end{eqnarray}
All the above coefficients are obtained up to one-loop order in the
electroweak interactions. The momentum $q$ is the momentum carried by
the virtual gluon in the penguin diagram. Note that $c_4^i = c_6^i$,
$i=u,c$, while $c_4^t \ne c_6^t$. This will be important in what
follows.

For the decay $B \to f_1 f_2$, we are really interested in the matrix
elements of the various operators. We therefore define new
coefficients ${\bar{c}}_i^{u,c}$ as
\bea
{\bar{c}}_i^{u,c} & = 
& \frac{\bra{f_1 f_2}c_i^{u,c}(q^2)O_i\ket{B}}
 {\bra{f_1 f_2}O_i\ket{B}} ~.
\eea
{}From Eq.~(\ref{coeffs}), the ${\bar{c}}_i^{u,c}$ can be expressed in
terms of the function $\bar{G}_i^{u,c}$ defined as
\bea 
{\bar{G}}_i^{u,c}(m_{u,c},\mu) & = & \frac{\bra{f_1
    f_2}G(m_{u,c},\mu,q^2)O_i\ket{B}} {\bra{f_1 f_2}O_i\ket{B}} ~.
\label{Gbar}
\eea
One can use models to calculate the functions $\bar{G}_i^{u,c}$, and 
one finds in general that the
functions $\bar{G}_i^{u,c}$ are process dependent. More importantly,
it is a reasonable assumption that the functions $\bar{G}_i^{u,c}$ are
independent of $i$: 
\beq
{\bar{G}}_i^{u,c}(m_{u,c},\mu) = {\bar{G}}_j^{u,c}(m_{u,c},\mu) ~.
\label{Gbarrel}
\eeq
This is because the effects of the different Dirac structures of the
operators $O_i$ cancel in the ratio in Eq.~(\ref{Gbar})
\cite{Kamal:1997wt}. This implies that the relations between the various
${\bar{c}}_i^{u,c}$ are the same as those between the various
$c_i^{u,c}(q^2)$. In particular, we have
\beq
{\bar c}_4^i = {\bar c}_6^i ~, i=u,c ~.
\label{cbarrel}
\eeq
This will be the key ingredient in the analysis below. In the study of
non-leptonic decays the usual practice is to replace
\bea
{\bar{c}}_i^{u,c} & \to & c_i^{u,c}(q^2_{av}) ~,
\eea
where $q^2_{av}$ is allowed to vary between $m_b^2/4$ and $m_b^2/2$ to
account for process dependence \cite{Deshpande:1990kc,Simma:1991ct}.

The structure of the effective Hamiltonian allows us to write the
amplitude for $\bsbar \to \phi \ks$ as
\beq
A_s^{\phi\ks} \equiv A(\bsbar \to \phi \ks) = {G_F \over \protect
  \sqrt{2}} (V_{ub}V^*_{ud}P_u +V_{cb}V^*_{cd}P_c +V_{tb}V^*_{td}P_t)
~.
\label{BsphiKsamp}
\eeq
Under the assumption of naive factorization one can write, dropping
factors common to $P_{u,c,t}$, and using the fact that $\bar{c}_6^i=
\bar{c}_4^i ~, i=u,c$ [Eq.~(\ref{cbarrel})],
\beq
P_{u,c} = \bar{c}_6^{u,c} (1-\frac{1}{N_c^2})
\left[\left\langle{O_{LL}}\right\rangle
-2\left\langle{O_{SP}}\right\rangle \right] ~,
\label{Puceqn}
\eeq
where
\bea
\left\langle{O_{LL}}\right\rangle
& = &
\bra{\phi}\bar{s}\gamma_{\mu}(1-\gamma_5)b\ket{\bsbar}
\bra{\ks}\bar{d}\gamma^{\mu}(1-\gamma_5)s\ket{0} ~, \nn\\
\left\langle{O_{SP}}\right\rangle & = &
\bra{\phi}\bar{s}(1-\gamma_5)b\ket{\bsbar}
\bra{\ks}\bar{d}(1+\gamma_5)s\ket{0} ~.
\label{LLSPdefs}
\eea
(The operator $O_{SP}$ appears due to a Fierz transformation: $(V-A)
\otimes (V+A) = -2 (S-P) \otimes (S+P)$.) On the other hand, the
contribution from the top penguin is more complicated:
\bea
P_t & = &\left[(c_4^t +\frac{c_3^t}{N_c})\left\langle{O_{LL}}\right\rangle
+(c_3^t +\frac{c_4^t}{N_c})\left\langle{O_{LL1}}\right\rangle\right] 
\nn\\
& + &
\left[-2(c_6^t +\frac{c_5^t}{N_c})\left\langle{O_{SP}}\right\rangle
+(c_5^t +\frac{c_6^t}{N_c})\left\langle{O_{LR1}}\right\rangle\right]
\nn\\
& - &\frac{1}{2}
\left[(c_9^t +\frac{c_{10}^t}{N_c})\left\langle{O_{LL1}}\right\rangle
+(c_{10}^t +\frac{c_9^t}{N_c})\left\langle{O_{LL}}\right\rangle\right] ~,
\eea
where
\bea
\left\langle{O_{LL1}}\right\rangle
& = &
\bra{\phi}\bar{s}\gamma_{\mu}(1-\gamma_5)s\ket{0}
\bra{\ks}\bar{d}\gamma^{\mu}(1-\gamma_5)b\ket{\bsbar} ~, \nn\\ 
\left\langle{O_{LR1}}\right\rangle
& = &
\bra{\phi}\bar{s}\gamma_{\mu}(1+\gamma_5)s\ket{0}
\bra{\ks}\bar{d}\gamma^{\mu}(1-\gamma_5)b\ket{\bsbar} ~.
\label{tquark1}
\eea
In the above, we have neglected the contributions from $c_{7,8}$. 

It is convienient to rewrite $P_{u,c}$ and $P_t$ as
\bea
P_{u,c} & = & \bar{c}_6^{u,c} (1-\frac{1}{N_c^2})X
\left\langle{O_{LL}}\right\rangle ~, \nn\\
P_t & = & a_6X\left\langle{O_{LL}}\right\rangle
+ (a_4-a_6 -\frac{1}{2}a_{10})\left\langle{O_{LL}}\right\rangle + 
(a_3+a_5 -\frac{1}{2}a_{9})\left\langle{O_{LL1}}\right\rangle ~,
\label{tquark2}
\eea
where $\left\langle{O_{LL1}}\right\rangle
=\left\langle{O_{LR1}}\right\rangle$,
\beq
a_i = \cases{ c_i + {c_{i-1} \over N_c} ~, & $i = 4,6,10$ ~,\cr
              c_i + {c_{i+1} \over N_c} ~, & $i = 3,5,9$ ~,\cr}
\eeq
and
\beq
X \equiv \left[ 1 - {2 \left\langle{O_{SP}}\right\rangle \over
    \left\langle{O_{LL}}\right\rangle} \right] ~.
\eeq
It is this latter quantity $X$ which is the focus of our attention in
this paper.

Using the fact that
\beq 
\langle \ks (p_{\sss K})|{\bar d}\gamma_{\mu}(1-\gamma_5) s|~0\rangle
= i f_{K_{S}}p_{{\sss K}\mu} ~,
\eeq
along with the equations of motion for the quarks (we assume that
$p_{\sss K} = p_d + p_{\bar s}$), it is straightforward to show that
\beq 
X = \left[ 1 - 2 \, {1 \over m_b + m_s} \, {m_{\sss K}^2 \over m_s +
m_d} \right] ~.  
\label{Xdef}
\eeq
However, the key point is the following: taking $\mK = 500$ MeV, $m_b
= 4.9$ GeV, $m_s = 100$ MeV (all at the $b$-quark mass scale), and
$m_d \simeq 0$, one finds that $X=0$!  Thus, the matrix elements
vanish for $u$ and $c$ but do {\it not} vanish for $t$. The decay
$\bs(t) \to \phi\ks$ is therefore dominated by a single decay
amplitude --- the $t$-quark penguin contribution --- and a measurement
of the CP-violating rate asymmetry probes the angle $\beta$
[Eq.~(\ref{Bsbtod})].

This result is related specifically to two facts: (i) the decay is a
pure penguin decay, and (ii) the final state $\phi\ks$ consists of a
vector and a pseudoscalar meson. This can be seen as follows. First,
consider a decay such as $\bd(t) \to \pi^+ \pi^-$, which receives a
tree-level contribution $T V_{ub} V_{ud}^* {\cal O}_{\sss T}$, where
the operator ${\cal O}_{\sss T}$ is of the form $(V-A) \otimes (V-A)$.
As we saw above, the matrix element of this operator alone does not
vanish. Thus, one must consider decays which have no tree-level
contributions. That is, only processes involving the quark-level
transition $b\to d s{\bar s}$ need be considered. Now consider a
pure-penguin final state containing two pseudoscalars, such as $\bd(t)
\to K^0 \kbar$. Repeating the above analysis, one finds that
\beq 
X = \left[ 1 + 2 \, {1 \over m_b - m_s} \, {m_{\sss K}^2 \over m_s +
m_d} \right] ~.
\eeq
Clearly there is now no cancellation between the two contributions,
and one finds $X\ne 0$. Similarly, if the final state consisted of two
vector mesons, such as in $\bd(t) \to K^{*0} {\bar K}^{*0}$, then the
$\left\langle{O_{SP}}\right\rangle$ matrix element would vanish due to
conservation of angular momentum, and again we would find $X\ne 0$.
Thus, only final states which consist of a vector and a pseudoscalar
can have $X=0$.

In fact, $\bs(t) \to \phi\ks$ is the {\it only} decay involving a
$b\to d$ penguin amplitude for which the ${\cal O}_u$ and ${\cal O}_c$
matrix elements vanish. The only other $b\to d s{\bar s}$ decay whose
final state consists of a vector and a pseudoscalar is $\bd(t) \to
\kbar K^{0*}$. However, $\bd\to \kbar K^{0*}$ and $\bdbar \to \kbar
K^{0*}$ do not factorize in the same way. Specifically, for $\bd\to
\kbar K^{0*}$, the $\kbar$ is coupled to the vacuum as in
Eq.~(\ref{LLSPdefs}), so that $X=0$. However, for $\bdbar \to \kbar
K^{0*}$, it is the $K^{0*}$ which is coupled to the vacuum, which
implies that the $\left\langle{O_{SP}}\right\rangle$ matrix element
vanishes. Thus, $X\ne 0$ for this decay. Therefore it is only the
decay $\bs(t) \to \phi\ks$ whose indirect CP asymmetry is expected to
measure $\sin 2\beta$.

Now, the vanishing of the $u$- and $c$-quark matrix elements in
$\bs(t) \to \phi\ks$ depends on two theoretical ingredients: (i) we
have assumed that naive factorization is valid for this decay, and
(ii) we have taken particular values for $m_s$ and $m_b$. It is
important to examine the extent to which these assumptions are
justified, and to show how the result changes when these assumptions
are relaxed.

First of all, we can see the effect of a nonzero value of $X$ as
follows. The measurement of the time-dependent rate $\bs(t) \to
\phi\ks$ allows one to extract both direct and indirect CP-violating
asymmetries. These are defined as follows:
\bea
a^{\sss CP}_{dir} & = & 
{ |A_s^{\phi\ks}|^2 - |{\bar A}_s^{\phi\ks}|^2 \over
 |A_s^{\phi\ks}|^2 + |{\bar A}_s^{\phi\ks}|^2} ~, \nn\\
a^{\sss CP}_{indir} & = & { 
{\rm Im} \left( {A_s^{\phi\ks}}^* {\bar A}_s^{\phi\ks} \right) \over
 |A_s^{\phi\ks}|^2 + |{\bar A}_s^{\phi\ks}|^2} ~,
\eea
where we have assumed that the weak phase of $\bs$--$\bsbar$ mixing is
negligible, as is the case in the SM. $A_s^{\phi\ks}$ is defined in
Eq.~(\ref{BsphiKsamp}), and ${\bar A}_s^{\phi\ks}$ is obtained from
$A_s^{\phi\ks}$ by changing the signs of the weak phases. Using CKM
unitarity to eliminate the $V_{ub}V^*_{ud}$ term in
Eq.~(\ref{BsphiKsamp}), $A_s^{\phi\ks}$ can be written as
\beq
A_s^{\phi\ks} = {G_F \over \protect \sqrt{2}}
( {\cal P}_{cu} e^{i\delta_c} + 
{\cal P}_{tu} e^{i\delta_t} e^{-i \beta} ) ~,
\eeq
where we have explicitly separated out the strong phases $\delta_c$
and $\delta_t$, as well as the weak phase $\beta$. The magnitudes of
the CKM matrix elements have been absorbed into the definitions of
${\cal P}_{cu}$ and ${\cal P}_{tu}$. Using this expression for
$A_s^{\phi\ks}$, the CP asymmetries take the form
\bea
a^{\sss CP}_{dir} & = & 
{ 2 {\cal P}_{cu} {\cal P}_{tu} \sin\beta \sin\Delta \over
{\cal P}_{tu}^2 + {\cal P}_{cu}^2 + 2{\cal P}_{tu} {\cal P}_{cu}
\cos\beta \cos\Delta} ~, \nn\\
a^{\sss CP}_{indir} & = & 
{ {\cal P}_{tu}^2 \sin 2\beta +
2 {\cal P}_{cu} {\cal P}_{tu} \sin\beta \cos\Delta \over
{\cal P}_{tu}^2 + {\cal P}_{cu}^2 +2 {\cal P}_{tu} {\cal P}_{cu}
\cos\beta \cos\Delta} ~,
\label{CPasymmetries}
\eea
where $\Delta \equiv \delta_t - \delta_c$. From these expressions, we
see that a nonzero value of $X$ corresponds to a nonzero value of
${\cal P}_{cu}$. This in turn leads to a nonzero value of the direct
CP asymmetry $a^{\sss CP}_{dir}$, and also affects the clean
extraction of $\sin 2\beta$ from the indirect CP asymmetry. In order
to compute the error on the measurement of $\sin 2\beta$, we will need
to estimate the size of the ratio ${\cal P}_{cu} / {\cal P}_{tu}$, as
well as the strong phase $\Delta$.

The main factor which may contribute to $X\ne 0$ is if the masses
$m_b$ and $m_s$ do not take the respective values 4.9 GeV and 100 MeV.
In computing the CP asymmetries in Eq.~(\ref{CPasymmetries}), we will
allow $m_b$ and $m_s$ to take a range of values. In our calculation,
we use current quark masses, evaluated at the scale $\mu \sim m_b$.
For the $b$-quark mass, we take $4.35 \le m_b \le 4.95$ GeV. As for
the current strange-quark mass, there is a great deal of uncertainty
in the value of $m_s$. One can obtain $m_s(\mu=m_b)$ by using as input
$m_s(\mu=1$ GeV) and then using QCD running to increase the scale up
to $\mu=m_b$. However, $m_s(\mu=1$ GeV) is not very well known and can
range from 0.150 to 0.200 GeV \cite{Fusaoka:1998vc}. In addition, it
is not clear that perturbative calculations are reliable near $\mu
\sim 1$ GeV. Given all these uncertainties, we vary $m_s(m_b)$ in the
range $0.08 \le m_s \le 0.12$ GeV.

There are also nonfactorizable effects which might give rise to $X\ne
0$. There have been several attempts to calculate corrections to the
naive factorization assumption. One promising approach is QCD-improved
factorization \cite{Beneke:1999br}, in which one systematically
calculates corrections to naive factorization in an expansion in
$\alpha_s(m_b) \sim 0.2$ and $\Lambda_{QCD}/m_b$. Naive factorization
appears as the leading-order term in this expansion. If we consider
QCD corrections to this term, we note that the $P_{u,c}$ arise already
at $O(\alpha_s)$, and so they receive no corrections at this order. In
fact, the $P_{u,c}$ are part of the $O(\alpha_s)$ corrections to the
naive factorization results. There are additional $\alpha_s$
corrections which can be taken into account by the replacement $a_i
\to a_i^{eff} =a_i(1+r_i)$ in Eq.~(\ref{tquark2}), where $r_i \sim
O(\alpha_s)$ are process-dependent corrections to the naive
factorization assumption.

Corrections to the $P_{u,c}$ in Eq.~(\ref{Puceqn}) may then arise if
Eq.~(\ref{Gbarrel}) does not hold. One can then write
\beq
P_{u,c} = (1-\frac{1}{N_c^2})
\left[\bar{c}_4^{u,c}\left\langle{O_{LL}}\right\rangle
  -2\bar{c}_6^{u,c}\left\langle{O_{SP}}\right\rangle \right] =
\bar{c}_6^{u,c} (1-\frac{1}{N_c^2})\left[X +X_1\right]
\left\langle{O_{LL}}\right\rangle ~,
\eeq
where
\beq
X_1 =\frac{ \bar{c}_4^{u,c}- \bar{c}_6^{u,c}}{ \bar{c}_6^{u,c}} ~.
\eeq
Note that $X_1$ is complex, so that if one writes $X_1=|X_1|
e^{i\theta}$, one obtains
\bea
|P_{u,c}| & = & |\bar{c}_6^{u,c}| (1-\frac{1}{N_c^2})
 \sqrt{X^2 +X_1^2 + 2|X||X_1| \cos {\theta}}
\left\langle{O_{LL}}\right\rangle\ ~.
\eea
In what follows we assume that the effect of $X_1$ is small enough to
be absorbed in the uncertainty in the quark masses in the expression
for $X$ in Eq.~(\ref{Xdef}). In other words the effect of $X_1$ is
essentially taken into account by varying the quark masses in $X$.

In order to test the robustness of the claim that the indirect CP
asymmetry in $\bd(t) \to \phi\ks$ measures $\sin 2\beta$, we perform
the following analysis. We scan the entire parameter space,
calculating the CP asymmetries $a^{\sss CP}_{dir}$ and $a^{\sss
  CP}_{indir}$ of Eq.~(\ref{CPasymmetries}) at each point in this
space. We are especially interested in the quantity $\delta$, which
measures the fractional difference between the indirect CP asymmetry
and the true value of $\sin 2\beta$:
\beq
\delta \equiv { a^{\sss CP}_{indir} - \sin{2\beta} \over \sin{2\beta}
  }~.
\label{deltadef}
\eeq
In particular, we wish to compute what fraction of the parameter space
leads to a given value for $\delta$. This will give us some sense of
the extent to which the asymmetry in $\bd(t) \to \phi\ks$ truly probes
$\beta$.

In addition to the masses $m_b$ and $m_s$, the CP asymmetries depend
on other quantities. First, they depend on the momentum transfer
$q^2_{av}$, which we vary between $m_b^2/4$ and $m_b^2/2$. Second,
they involve the CKM parameters $\rho$ and $\eta$, whose ranges are
taken from Ref.~\cite{Ali:2001hy}. Third, one also needs values for
the $d$- and $c$-quark masses. We take $m_d(\mu=m_b)= 6$ MeV and $1.1
\le m_c \le 1.4$ GeV \cite{PDG}. Finally, note that, while $P_{u,c}$
depend only on the matrix element $\left\langle{O_{LL}}\right\rangle$,
$P_t$ also depends on $\left\langle{O_{LL1}}\right\rangle$
[Eq.~(\ref{tquark2})]. Using factorization, one can write
\beq
r^{\prime} = \frac{\left\langle{O_{LL1}}\right\rangle}
  {\left\langle{O_{LL}}\right\rangle}  = 
\frac{f_\phi F_1^{\ks} (m_{\phi}^2)}{f_{\ks}A_0^{\phi}(m_K^2)} ~,
\eeq
where the various formfactors and decay constants above are defined by
\bea
\langle \ks (q)|{\bar d}\gamma_{\mu}(1-\gamma_5) u|~0\rangle
&=&i f_{\ks}q_{\mu} ~, \nn\\
\langle \phi(q,~\epsilon)~|{\bar s}\gamma_{\mu}\gamma_5
b|~\bdbar(p)\rangle & = & (M_B +M_{\phi})A_1^\phi
\left[\epsilon_{\mu}^*
  -\frac{\epsilon^*.(p-q)}{(p-q)^2}(p-q)_{\mu}\right] \nn\\
& & \quad -A_2^\phi \frac{\epsilon^*.(p-q)}{M_B+M_{\phi}} \left[
  (P_B+P_{\phi})_{\mu} - 
  \frac{M_B^2-M_{\phi}^2}{(p-q)^2}(p-q)_{\mu}\right] \nn\\
& & \quad + 
 2M_{\phi} A_0^\phi \frac{\epsilon^*.(p-q)}{(p-q)^2}(p-q)_{\mu} ~, \nn\\
\langle \ks (q)~|{\bar u}\gamma_\mu b|~\bdbar(p) \rangle &=&
 F_1^{\ks}\left[(p+q)^\mu-\frac{m^2_B-m^2_{\ks}}{(p-q)^2}(p-q)^\mu\right] \nn\\
& & \quad +F_0^{\ks}\, \frac{m^2_B-m^2_{\ks}}{(p-q)^2}(p-q)^\mu ~, \nn\\
\langle 0~|{\bar s}\gamma^\mu s|~\phi(q,~\epsilon)\rangle 
&=&M_{\phi} f_\phi \epsilon^{ \mu} ~.
\eea  
Using $f_{\phi}=0.237$ GeV \cite{Cheng:2000hv} and the various form
factors calculated using the light cone sum rules \cite{Ball:1998kk}
we obtain $r^{\prime}\approx $ 1.5.

\begin{figure}[htb] 
\centerline{\epsfysize 4.2 truein \epsfbox{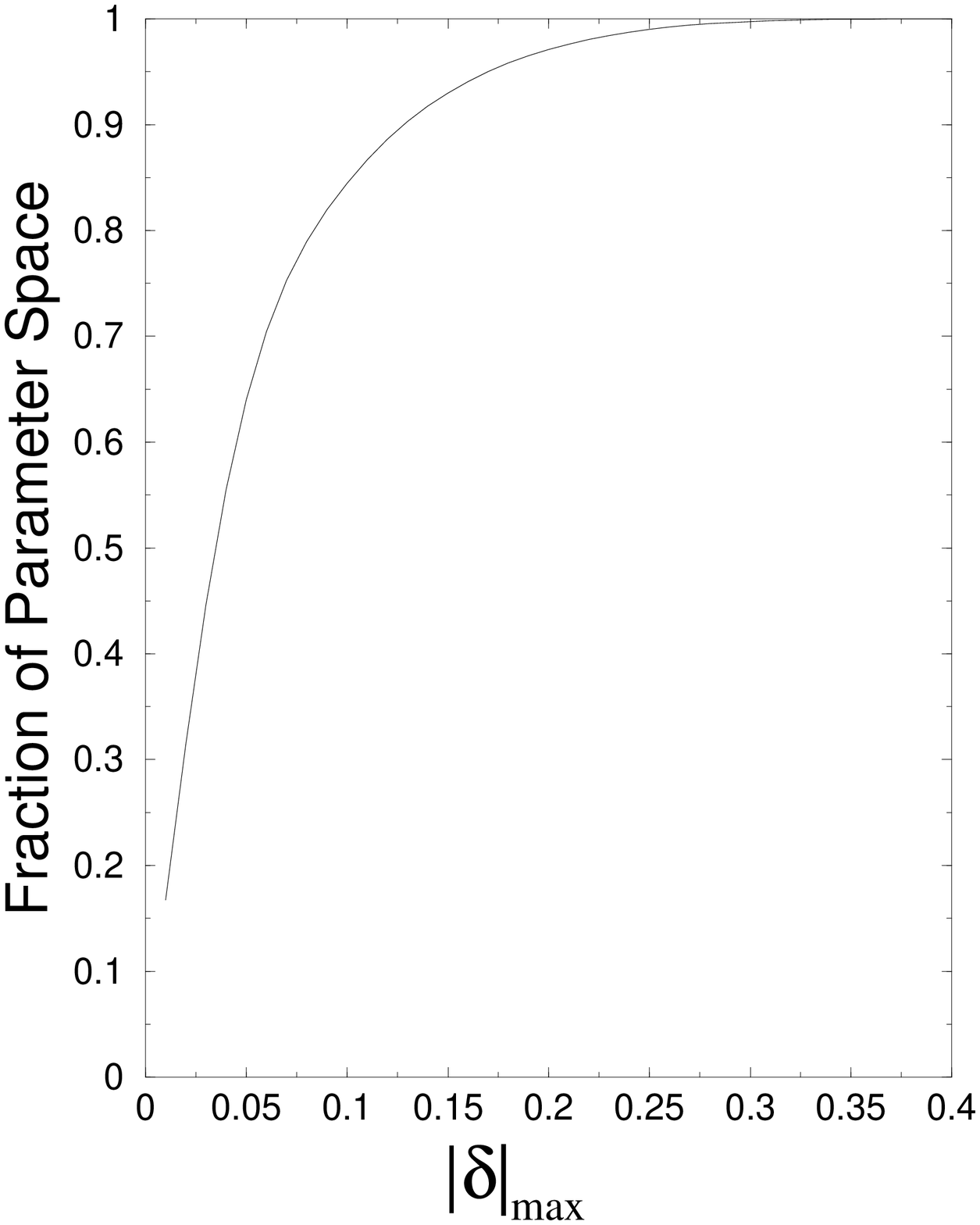}} 
\caption{The fraction of the parameter space for which the error 
  $|\delta|$ in extracting $\sin{2\beta}$ from $a^{\sss CP}_{indir}$
  is less than a given value, $|\delta|_{max}$.}
\label{delta} 
\end{figure}

We are now in a position to calculate the CP asymmetries $a^{\sss
  CP}_{dir}$ and $a^{\sss CP}_{indir}$ in $\bs(t)\to\phi\ks$
[Eq.~(\ref{CPasymmetries})], as well as the deviation of the indirect
CP asymmetry from the true value of $\sin 2\beta$. In
Fig.~\ref{delta}, we plot the fraction of the parameter space for
which $|\delta|$, the error on $\sin 2\beta$ as extracted from the
indirect CP asymmetry [Eq.~(\ref{deltadef})], is less than a given
value, $|\delta|_{max}$. From this figure, we see that $\sin 2\beta$
can be obtained with an error less than 30\% over virtually the entire
parameter space. And this error is reduced to about 10\% in 80\% of
the parameter space. While this should not be interpreted
statistically as some sort of confidence level, it does indicate that
it is quite likely that $\beta$ can be extracted from the indirect CP
asymmetry in $\bs(t)\to\phi\ks$ with a rather small error.

Of course, if the time-dependent rate for $\bs(t)\to\phi\ks$ is
measured, we will have more information than just $a^{\sss
  CP}_{indir}$: we will also measure the direct CP asymmetry $a^{\sss
  CP}_{dir}$. Since $a^{\sss CP}_{dir}$ vanishes if $X=0$, its value
may help us determine the extent to which $a^{\sss CP}_{indir}$ really
measures $\sin 2\beta$. At first glance, the correlation between
$a^{\sss CP}_{dir}$ and $a^{\sss CP}_{indir}$ appears airtight: if
$a^{\sss CP}_{dir}$ is found to vanish, then this must imply that
$X=0$, so that $a^{\sss CP}_{indir}$ yields $\sin 2\beta$.
Unfortunately, things are not quite so straightforward: $a^{\sss
  CP}_{dir}$ is also proportional to the strong phase difference
$\Delta$ [Eq.~(\ref{CPasymmetries})]. Therefore, if $\Delta \simeq 0$,
then $a^{\sss CP}_{dir}$ will vanish even if $X\ne 0$. Thus, this
possibility must be taken into account in evaluating the correlation
between the measurements of $a^{\sss CP}_{indir}$ and $a^{\sss
  CP}_{dir}$.

\begin{figure}[htb]  
\centerline{\epsfysize 4.2 truein \epsfbox{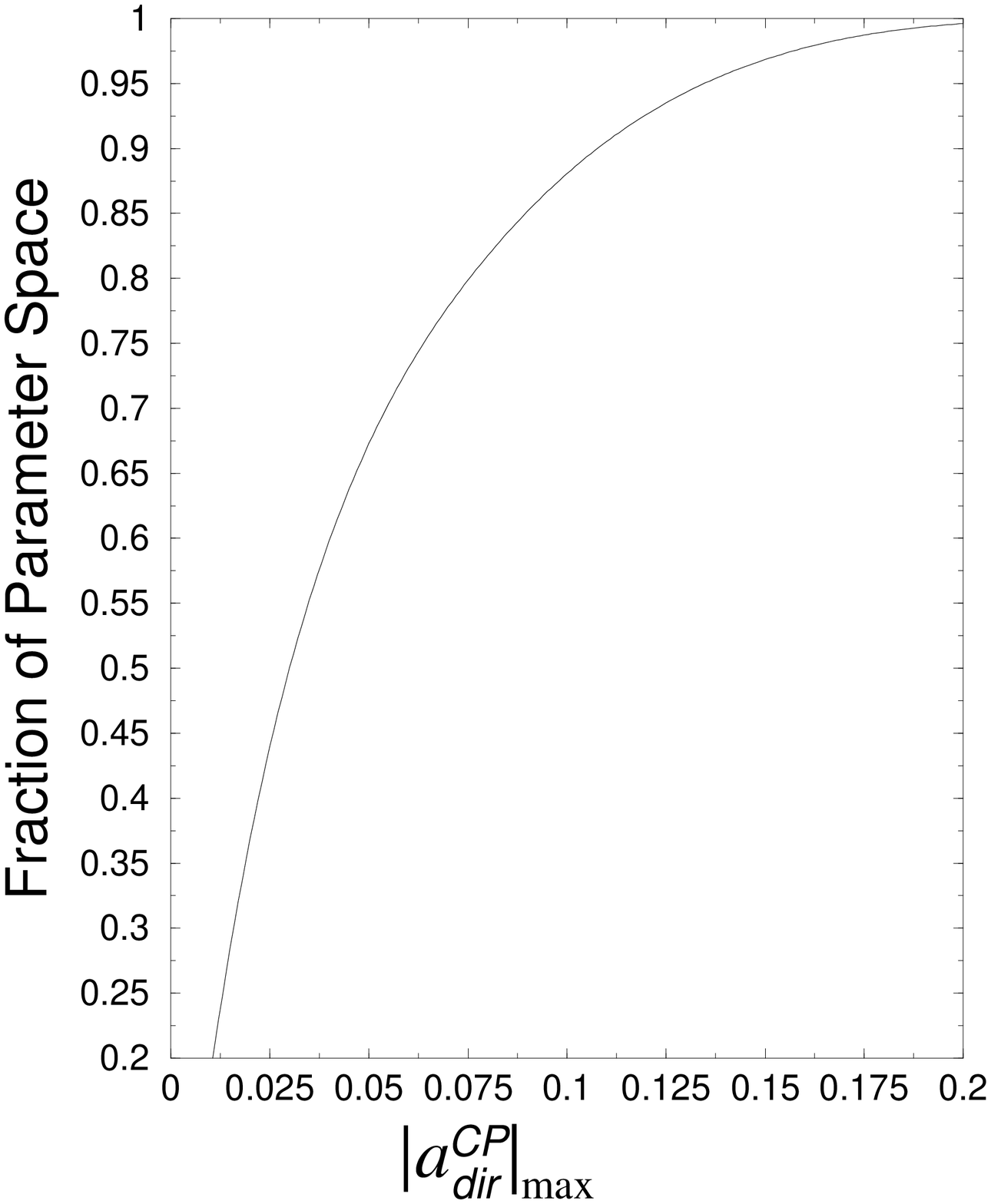}} 
\caption{The fraction of the parameter space for which 
  $a^{\sss CP}_{dir}$ is less than a given value, $|a^{\sss
    CP}_{dir}|_{max}$.}
\label{delta3} 
\end{figure} 

\begin{figure}[htb] 
\centerline{\epsfysize 4.2 truein \epsfbox{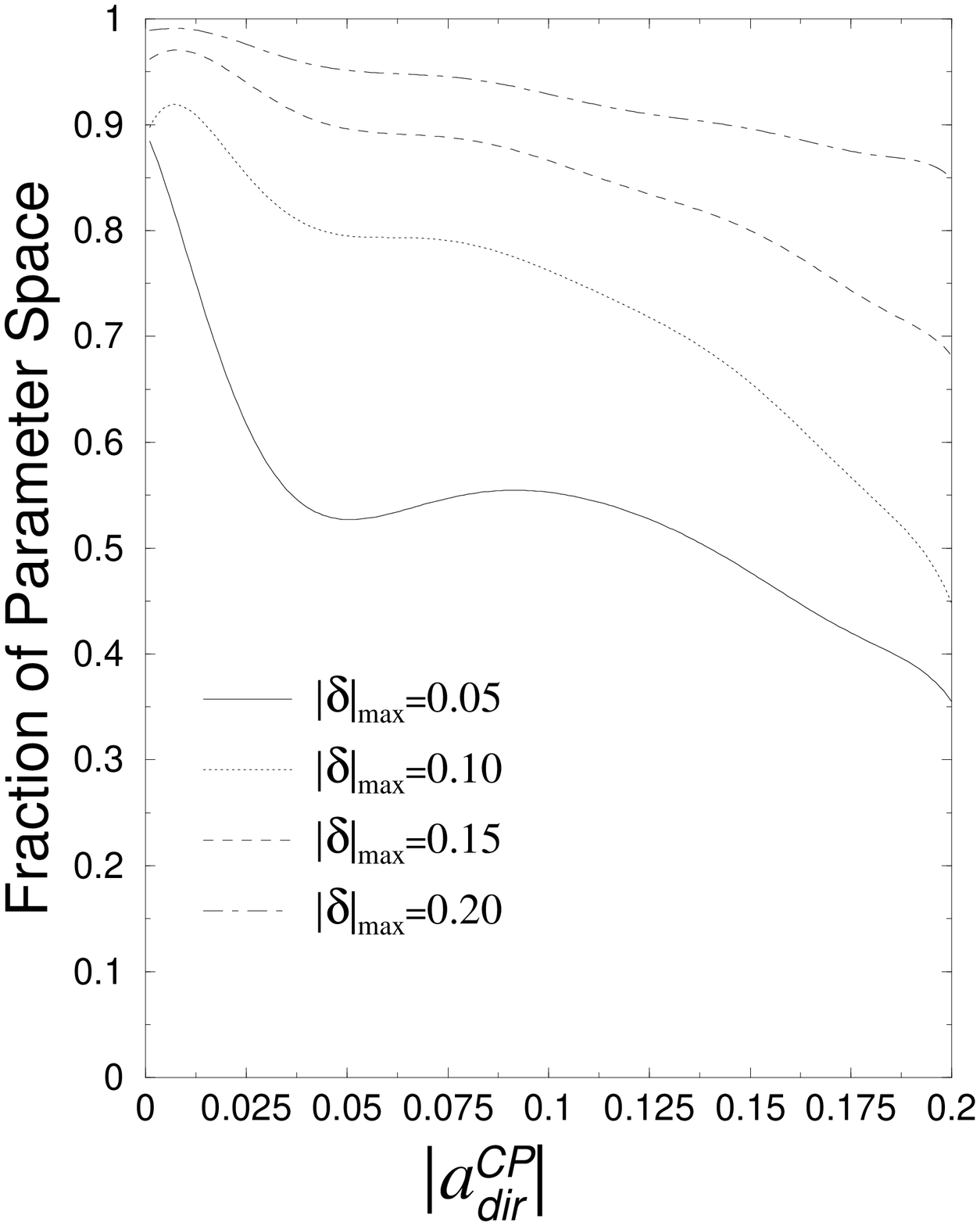}} 
\caption{The fraction of the parameter space for which
  $\delta$ can be measured with a maximum error of $|\delta|_{max} =$
  0.05, 0.1, 0.15 and 0.20, as a function of $|a^{\sss CP}_{dir}|$.}
\label{delta1} 
\end{figure}

In Fig.~\ref{delta3}, we plot the fraction of the parameter space for
which $a^{\sss CP}_{dir}$ is less than a given maximal value. As is
clear from the figure, we expect $a^{\sss CP}_{dir}$ to be at most
20\% -- larger values would point to the presence of new physics.
Fig.~\ref{delta1} shows, for a given value of $|a^{\sss CP}_{dir}|$,
the fraction of the parameter space for which $|\delta|$ is less than
a particular maximum error ($|\delta|_{max} =$ 5\%, 10\%, 15\%, 20\%).
{}From this plot, we see that if $a^{\sss CP}_{dir}$ is measured to be
0.1, one can obtain $\sin 2\beta$ from $a^{\sss CP}_{indir}$ with an
error of 5\% (20\%) in $\sim 55$\% ($\sim 95$\%) of the parameter
space. If $a^{\sss CP}_{dir}$ is found to be tiny, then this is
probably due to the fact that $X \simeq 0$, since $\delta < 5$\% over
$\sim 90$\% of the parameter space. However, as discussed above, this
does not hold over the entire space since $a^{\sss CP}_{dir}$ can be
small if $\Delta \simeq 0$, while $X\ne 0$.

There is one technical point which is worth mentioning here. Naively,
one would expect that, as $a^{\sss CP}_{dir}$ increases, the error
$\delta$ will also increase. That is, one expects that the fraction of
the parameter space with a given maximum error $|\delta|_{max}$ should
decrease with increasing $a^{\sss CP}_{dir}$. This is because the
prediction that the indirect CP asymmetry in $\bs(t)\to\phi\ks$
measures $\sin 2\beta$ depends on the fact that $X=0$, and $a^{\sss
  CP}_{dir}$ increases with increasing $X$. However, Fig.~\ref{delta1}
does not behave in exactly this way: although the fraction of the
parameter space with a given $|\delta|_{max}$ does indeed roughly
decrease with increasing $a^{\sss CP}_{dir}$, this decrease is not
monotonic. For example, the curve for $|\delta|_{max} = 0.05$ flattens
out at around $|a^{\sss CP}_{dir}| \sim 0.04$, and even turns up
before falling again for $|a^{\sss CP}_{dir}| \gsim 0.1$. The
explanation for this behaviour is as follows. From
Eq.~(\ref{CPasymmetries}), we see that $|a^{\sss CP}_{dir}|$ depends,
among other things, on the strong phase difference $\Delta$ and the
ratio $r={\cal P}_{cu}/{\cal P}_{tu}$. As $q^2_{av}$ is varied,
keeping other parameters fixed, we find that, as $r$ increases, so
does $\sin{\Delta}$. However, $|\delta|$ depends on $\cos{\Delta}$,
which obviously decreases as $\sin\Delta$ increases.  Thus, there is a
region of parameter space where, as $r$ and $\sin\Delta$ increase, the
error $|\delta|$ actually decreases due to the presence of the
$\cos\Delta$ term. It is this effect which is responsible for the
flattening out and the slight rise in the curves in Fig.~\ref{delta1}.

\begin{figure}[htb] 
\centerline{\epsfysize 4.2 truein \epsfbox{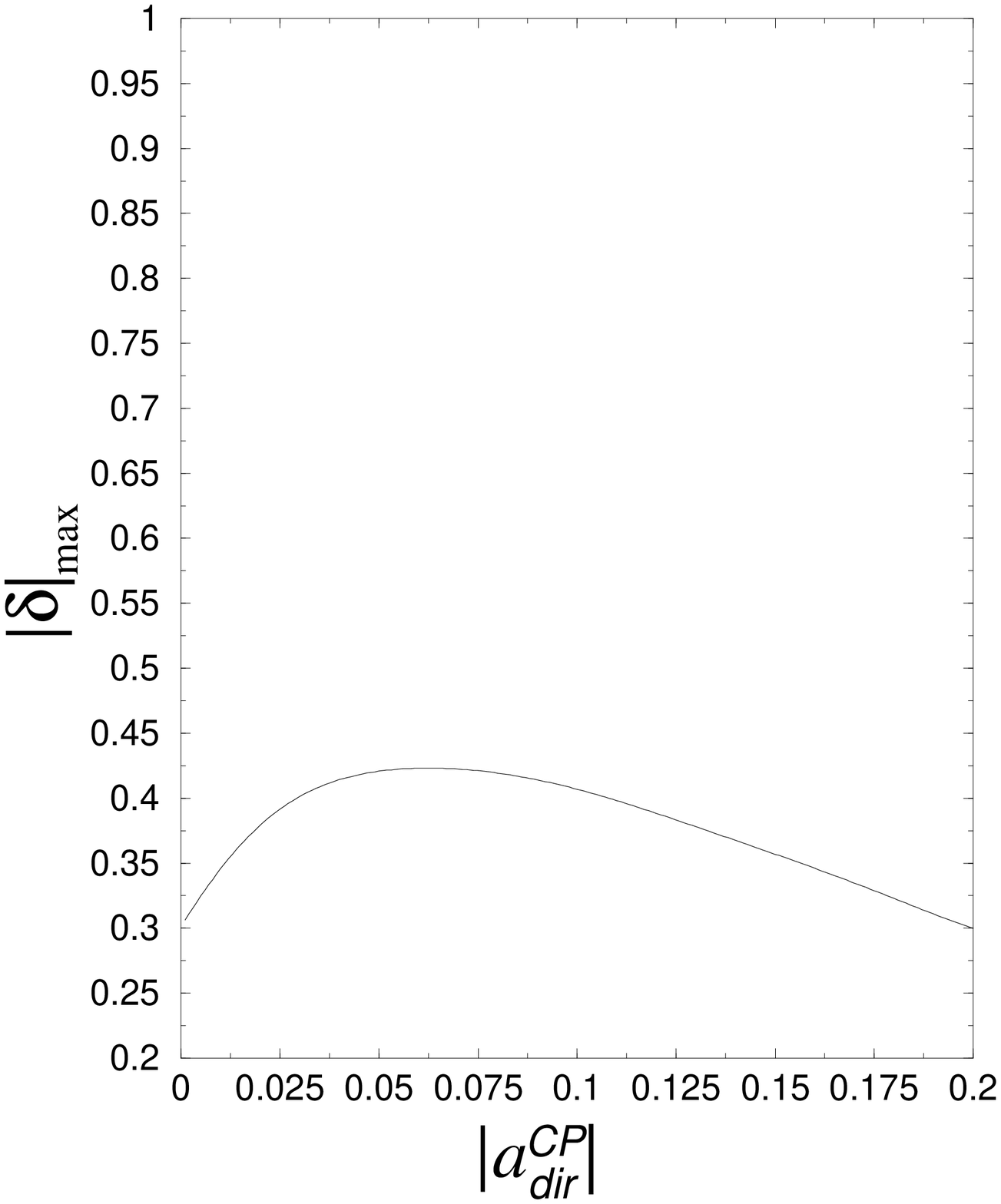}}
\caption{The maximum error $|\delta|_{max}$ for the entire parameter 
  space as a function of $|a^{\sss CP}_{dir}|$.}
\label{delta2} 
\end{figure} 

Finally, one can take a most conservative point of view, and ask what
is the maximum error $|\delta|_{max}$ over the entire parameter space
for a given measurement of $|a^{\sss CP}_{dir}|$. This is shown in
Fig.~\ref{delta2}. Depending on the value of $|a^{\sss CP}_{dir}|$,
$|\delta|_{max}$ is between 30 and 40\%. As was the case in
Fig.~\ref{delta1}, the downturn of the curve as $|a^{\sss CP}_{dir}|$
increases is due to the dependence of $\delta$ on $\cos\Delta$.

Of course, if $X$ does indeed vanish, this may have some negative
practical implications. Specifically, since there are fewer
contributions to the amplitude for $\bs\to\phi\ks$, one might suspect
that the branching ratio will be smaller than that of other pure $b\to
d$ penguin decays. This is indeed what we find: taking $\rho=0.18$ and
$\eta=0.36$ \cite{Ali:2001hy}, $P_u=P_c=$0, and using the form factors
in Ref~\cite{Ball:1998kk}, we obtain
\bea
\frac{BR[\bs \to \phi \ks]}{BR[\bdbar \to \rho^{+} \pi^{-}]} & =&
0.003 ~.
\eea
Using the measured $BR[\bdbar \to \rho^{+} \pi^{-}] = 30 \times
10^{-6}$ \cite{Jessop:2000bv} we find $ BR[\bs \to \phi \ks] \sim
10^{-7}$, which is very small.

Fortunately, the above analysis for $\bs(t)\to\phi\ks$ also applies to
the decay $\bs(t) \to \phi(1680)\ks$, where $\phi(1680)$ is a radially
excited $\phi$. We expect the branching ratio for $\bs \to
\phi(1680)\ks$ to be almost a factor 10 larger than $\bs \to \phi \ks$
\cite{Datta:2001hd}. This is because the form factor for $\bs \to
\phi$ (or, in general, for any $B \to$ light meson) probes the
high-momentum tail of the $\phi$ wavefunction. As the radially excited
$\phi(1680)$ has more high-momentum components, the form factor for
$\bs \to \phi(1680)$ is enhanced relative to $\bs \to \phi$.

Note that the measurement of CP violation in $\bd(t) \to \phi\ks$
probes $\beta$ [Eq.~(\ref{Bdbtos})]. If the measurement of $\beta$ as
extracted in this mode disagrees with the measurement of $\beta$ from
$\bd(t) \to J/\Psi\ks$, this will indicate the presence of new physics
in the $b\to s$ penguin amplitude, i.e.\ in the $b\to s$
flavour-changing neutral current (FCNC) \cite{LonSoni}. Similarly, the
value of $\beta$ extracted in $\bs(t)\to\phi\ks$ or $\bs(t) \to
\phi(1680)\ks$ can be compared with that found in $\bd(t) \to
J/\Psi\ks$. Assuming that the weak phase of $\bs$-$\bsbar$ mixing is
tiny --- and this can be tested by looking for CP violation in $\bs(t)
\to J/\Psi\phi$, for example --- a discrepancy between these two
values points clearly to new physics in the $b\to d$ FCNC. This new
physics might affect $\bd$-$\bdbar$ mixing and/or the $b\to d$ penguin
amplitude. Now, it is quite likely that CP violation in $\bs$ decays
can only be measured at hadron colliders, since one needs an extremely
large boost in order to resolve the rapid $\bs$-$\bsbar$ oscillations.
Since hadron colliders produce copious amounts of $\bd$ and $\bs$
mesons, it should be possible to perform the $\bd(t) \to \phi\ks$ and
$\bs(t) \to \phi\ks$ analyses simultaneously, since the final state is
the same. Thus, by measuring $\beta$ in these decay modes, at hadron
colliders one can test for the presence of new physics in both the
$b\to s$ and $b\to d$ FCNC's.

To summarize: in general, $b\to d$ penguin decays receive
contributions from internal $u$, $c$ and $t$ quarks. Because of this,
one cannot cleanly extract information about weak phases from
measurements of CP-violating rate asymmetries. In this paper, we have
shown that this does not hold for the decay $\bs(t)\to\phi\ks$. Due to
a fortuitous cancellation between the $(V-A) \otimes (V-A)$ and $(V-A)
\otimes (V+A)$ matrix elements, the contributions from the $u$- and
$c$-quark operators each vanish. Thus, $\bs(t)\to\phi\ks$ is dominated
by a single decay amplitude, that of the internal $t$-quark, so that
the indirect CP asymmetry measures $\sin 2\beta$. This is the only
$b\to d$ penguin decay for which this occurs.

Of course, this cancellation is not rigourous: it depends on our
choosing particular values for the $s$- and $b$-quark masses. However,
we have shown that, over most ($\sim 80$\%) of the theoretical
parameter space, the difference between the indirect CP asymmetry and
the true value of $\sin 2\beta$ is less than 10\%. Furthermore, by
using information about the direct CP asymmetry, one can get a better
handle on the probable error on $\sin 2\beta$. For example, should
$a^{\sss CP}_{dir}$ be found to be tiny, this increases the likelihood
that $\sin 2\beta$ is being extracted with a small error: for $a^{\sss
  CP}_{dir} \simeq 0$, the error on $\sin 2\beta$ is at most 5\% over
almost 90\% of the parameter space.

Because the $u$- and $c$-quark contributions to $\bs(t)\to\phi\ks$ are
very small, one can expect that the branching ratio for this decay
will be reduced. This is indeed what is found: we estimate that
$BR[\bs \to \phi \ks] \sim 10^{-7}$. However, the same analysis also
applies to the decay $\bs(t) \to \phi(1680)\ks$, where $\phi(1680)$ is
a radially excited $\phi$. The branching ratio for this decay is
expected to be about 10 times larger than that for $\bs(t)\to\phi\ks$.

Finally, we note that $\bs(t)\to\phi\ks$ will probably be studied at a
hadron collider. These same experiments can simultaneously measure the
CP asymmetry in $\bd(t)\to\phi\ks$, which also probes $\sin 2\beta$ in
the SM. By comparing the CP asymmetries in these two modes with that
measured in $\bd(t)\to J/\psi\ks$, one can look for the presence of
new physics in both $b\to d$ and $b\to s$ transitions.

\section*{\bf Acknowledgments}

C.S.K. thanks D.L. for the hospitality of the Universit\'e de
Montr\'eal, where some of this work was done. The work of C.S.K. was
supported in part by the BK21 Program, SRC Program and Grant No.
2000-1-11100-003-1 of the KOSEF, and in part by the KRF Grants,
Project No. 2000-015-DP0077. The work of A.D. and D.L. was
financially supported by NSERC of Canada.

\end{document}